\documentclass[sigconf, nonacm]{acmart}

\acmConference[2024 Google PhD Fellowship Research Proposal]{2024 Google PhD Fellowship Research Proposal}

\begin{document}

\title[Towards Intent-based User Interfaces: Charting the Design Space of Intent-AI Interactions Across Task Types
]{Towards Intent-based User Interfaces: Charting the Design Space of Intent-AI Interactions Across Task Types}

\author{Zijian Ding}
\email{ding@umd.edu}
\orcid{0000-0002-6372-0369}
\affiliation{%
  \institution{University of Maryland, College Park}
  \country{USA}
}

\begin{abstract}
Technological advances continue to redefine the dynamics of human-machine interactions, particularly in task execution. This proposal responds to the advancements in Generative AI by outlining a research plan that probes intent-AI interaction across a diverse set of tasks: fixed-scope content curation task, atomic creative tasks, and complex and interdependent tasks. This exploration aims to inform and contribute to the development of Intent-based User Interface (IUI). The study is structured in three phases: examining fixed-scope tasks through news headline generation, exploring atomic creative tasks via analogy generation, and delving into complex tasks through exploratory visual data analysis. Future work will focus on improving IUIs to better provide suggestions to encourage experienced users to express broad and exploratory intents, and detailed and structured guidance for novice users to iterate on analysis intents for high quality outputs.
\end{abstract}

\newcommand{\todo}[1]{{\color{blue} #1}}

\begin{teaserfigure}
\centering \includegraphics[width=0.63\textwidth]{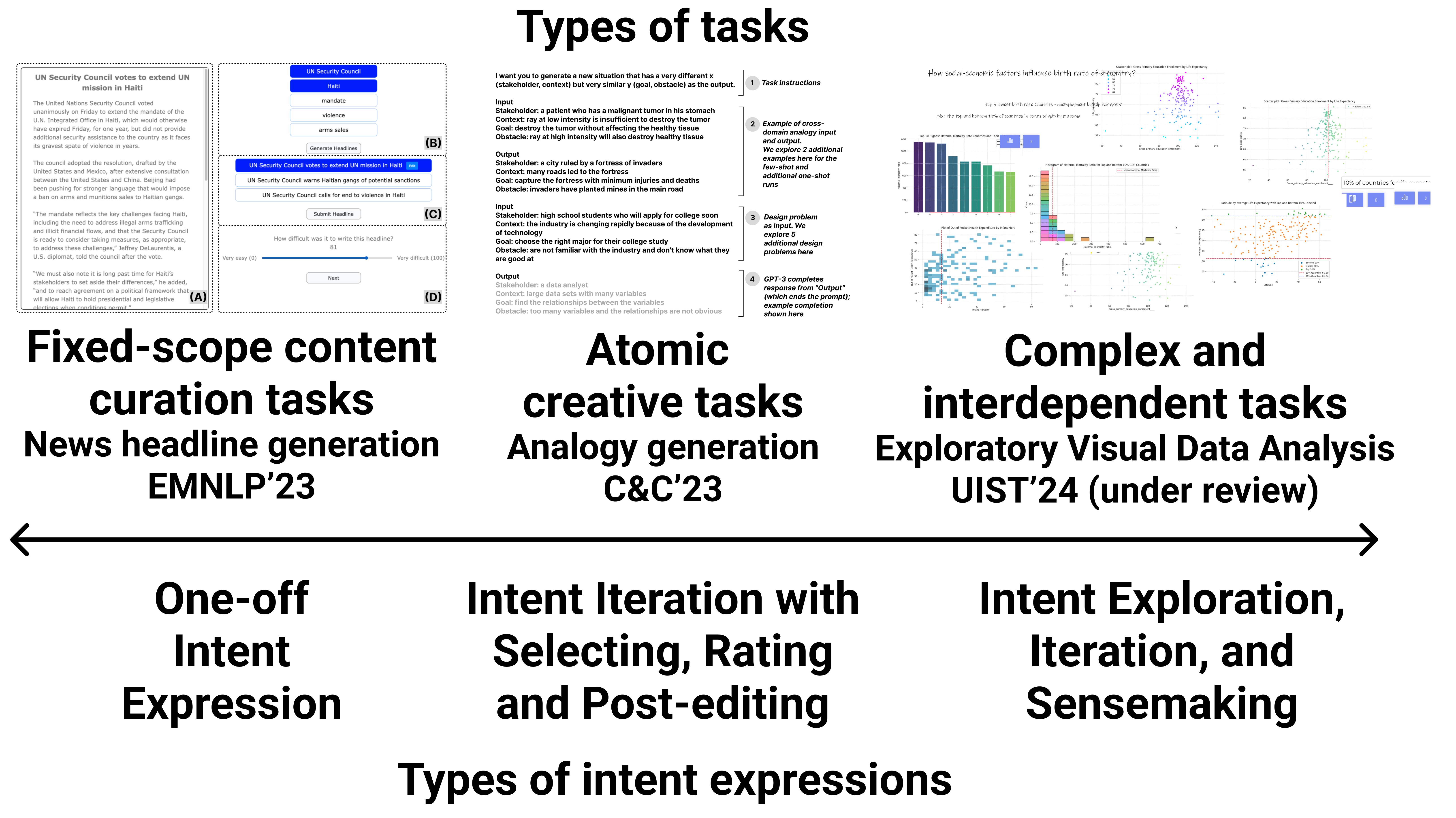}
\caption{The spectrum of human intent expression ranges from one-off intent expression for fixed-scope content curation tasks \cite{dingHarnessingPowerLLMs2023}, to intent iteration with selecting, rating and post-editing for atomic creative tasks \cite{ding2023fluid}, and intent exploration, iteration and sensemaking for complex and interdependent tasks \cite{dingIntelligentCanvasEnabling2024}. The spectrum is based on human-AI interaction paradigms \cite{dingAdvancingGUIGenerative2024,dingMappingDesignSpace2023}.}
\label{fig:spectrum}
\Description{TODO}
\end{teaserfigure}

\maketitle

\section{Introduction}

Human-computer interaction (HCI) involves humans expressing intent for a task, which the machine then interprets, executes, and communicates the results back to the user. Historically, the capabilities of machines to interpret human intents and execute tasks were limited, resulting in complex and burdensome processes for expressing these intents. This complexity often necessitated the mastery of intricate programming languages for command-line interfaces or the manipulation of buttons, dropdown menus, and other elements within graphical user interfaces (GUIs). The advent of generative AI, particularly the emergence of large language models, enables humans — at least for simple tasks like text summarization — to convey their intents to machines using natural language prompts, which are then translated into actions. Building on these shifts in machine capabilities and interaction paradigms, we are inspired to explore the concept of Intent-based User Interfaces (IUIs). Here, "intent" refers to the overarching purpose or objective at a higher level in the causal chain of a task, rather than the specific steps such as direct manipulations on complex GUIs. For instance, in a data analysis programming environment, the primary intent is to understand trends in the data, as opposed to manually inputting column names in plotting functions. This leads us to our research question for the design space of IUI: \textit{What kinds of intent-AI interaction are needed across different types of tasks?}

To explore the design space of IUI, which bridges user intents with task execution, we have investigated the types of intent-based interactions required for various types of tasks.. Figure \ref{fig:spectrum} illustrates a spectrum of these intent-based interactions that extend from one-off expressions of intent for fixed-scope content curation tasks like news headline generation \cite{dingHarnessingPowerLLMs2023}, through intent iterations with selecting, rating  and post-editing for atomic creative tasks such as analogy-driven problem reformulation \cite{ding2023fluid}. Current work is exploring what intent-based interactions are needed for complex and interdependent tasks,, such as exploratory visual data analysis: here, we have some evidence that intent sensemaking with affordances like exploration iteration, and curation are needed to support intent-based interactions \cite{dingIntelligentCanvasEnabling2024}.

\section{Related work}

My doctoral research studied the paradigms of intent expression on generative AI across a spectrum of tasks, varying from simple to complex and interdependent endeavors. This spectrum encompasses fixed-scope content curation tasks, atomic creative tasks, and complex and interdependent tasks (refer to Figure \ref{fig:spectrum}).

\subsection{Fixed-scope Content Curation Tasks}

Substantial research demonstrates that Large Language Models (LLMs) adeptly manage defined content curation tasks such as text summarization \cite{goyalNewsSummarizationEvaluation2022}, content refinement \cite{linWhyHowEmbrace2023}, and code explanation \cite{yanIvieLightweightAnchored2024}. These tasks involve reiterating existing knowledge succinctly and coherently without generating novel insights. LLMs have shown proficiency in these tasks, often without the necessity for human intervention after expressing their intents of tasks. For example, Clark et al. revealed that text generated by LLM was linguistically advanced to the extent that distinguishing it from human-written text became challenging for evaluators \cite{clarkAllThatHuman2021}.

\subsection{Atomic Creative Tasks}

The second category comprises atomic creative tasks that necessitate generating novel and valuable outputs \cite{sawyerExplainingCreativityScience2012,chanFormulatingFixatingEffects2024}. These include tasks such as crafting analogous design concepts \cite{zhuGenerativePreTrainedTransformer2022, leeEvaluatingHumanLanguageModel2022}, design problems \cite{macneilProbMapAutomaticallyConstructing2021,macneilFindingPlaceDesign2021,macneilFramingCreativeWork2021}, slogans \cite{clarkCreativeWritingMachine2018}, and tweetorials \cite{geroSparksInspirationScience2022}. Large Language Models (LLMs) can foster novel connections or "creative leaps" due to their extensive knowledge base \cite{chanSemanticallyFarInspirations2017}. However, real creativity often requires domain-specific and nuanced knowledge that might be absent in LLM training data. Hence, to ensure the final artifacts fulfill the task requirement, LLM outputs for these tasks need to be iterated in the format of guiding, selecting, or post-editing.

\subsection{Complex and Interdependent Tasks}

Beyond the atomic creative tasks are complex and interdependent tasks such as active search \cite{palaniCoNotateSuggestingQueries2021,palaniActiveSearchHypothesis2021}, data visualization \cite{vaithilingamDynaVisDynamicallySynthesized2024}, video editing \cite{wangLAVELLMPoweredAgent2024}, active search \cite{palaniActiveSearchHypothesis2021,palaniCoNotateSuggestingQueries2021}, design workshops \cite{macneilFreeformTemplatesCombining2023}, neurocognitive disorders screening \cite{dingTalkTiveConversationalAgent2022}, and storytelling in text \cite{yuanWordcraftStoryWriting2022a, singhWhereHideStolen2022} and images \cite{yanXCreationGraphbasedCrossmodal2023}. These tasks demand not only domain expertise but also capabilities for planning, reasoning, ideating, and maintaining context over extended periods to create coherent and innovative content. Researchers have developed LLM-supported tools like LAVE \cite{wangLAVELLMPoweredAgent2024} for video editing, DynaVis \cite{vaithilingamDynaVisDynamicallySynthesized2024} for data visualization, and XCreation \cite{yanXCreationGraphbasedCrossmodal2023} for cross-modal storytelling to facilitate these intricate interactions.

\section{Study Plan and Preliminary Results}

Our research comprises three studies, each examining intent-AI interactions across tasks varying in creativity and complexity. The first study examines fixed-scope content curation tasks, employing news headline generation as the medium for exploration. The second study transitions to probing atomic creative tasks, with a focus on analogy generation to uncover underlying creative processes. Lastly, the third ongoing study explores the complex and interdependent tasks associated with exploratory visual data analysis. We hope to use these three studies to collectively offer a comprehensive understanding of intent-AI interaction across a spectrum of task.

\subsection{Study 1 on News Headline Generation: One-off Intent Expression for Fixed-Scope Content Curation Tasks}

\begin{figure}
\centering
\includegraphics[width=0.8\linewidth]{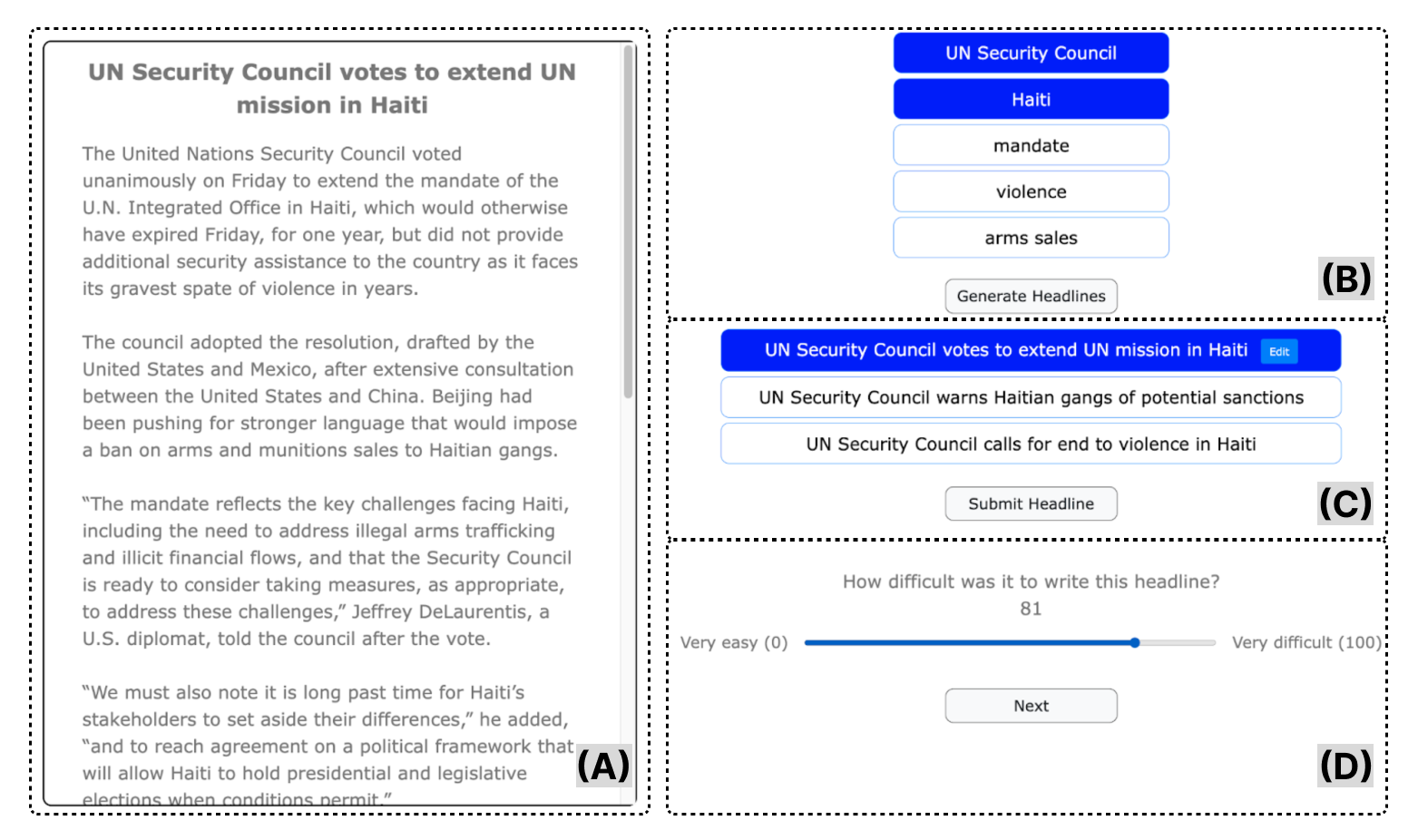}
\caption{Study 1 on fixed-scope content curation tasks - interface for human-AI news headline co-creation for \textit{guidance + selection + post-editing} condition: (A) news reading panel, (B) perspectives (keywords) selection panel (multiple keywords can be selected), (C) headline selection panel with post-editing capability, and (D) difficulty rating slider.}
\label{fig:study1_interface}
\end{figure}

This research explores intent-AI interactions within fixed-scope content curation tasks, aiming to inform the design of Intent-based User Interfaces (IUIs). We specifically examine the generation of news headlines from articles as a representative task of text summarization \cite{dingHarnessingPowerLLMs2023}. We conducted a between-subjects experiment involving the use of a large language model (LLM), GPT-3 \textit{text-davinci-002}, with 40 participants. These participants were tasked with creating news headlines under six conditions:

\begin{itemize}
\item \textit{Baseline}: Utilization of headlines currently employed by online media.
\item \textit{One-off intent expression}: Headlines generated by GPT-3.5 davinci-002 in response to the prompt: "Please generate three high-quality (attractive, clear, accurate, and inoffensive) headlines for the following news article."
\item \textit{Iteration with selection}: The LLM generates three headlines, from which the user selects the most fitting one.
\item \textit{Iteration with guidance and selection}: The LLM identifies potential keywords from each news article; the user selects one or more keywords to guide the headline generation process, followed by the selection of the most fitting headline.
\item \textit{Iteration with guidance, selection, and post-editing}: Similar to the previous condition, with an added feature allowing the user to edit the selected headline. An example of this interactive interface is presented in Figure \ref{fig:study1_interface}.
\item \textit{Manual only}: Participants write the headlines independently, without AI assistance.
\end{itemize}

The effectiveness of these conditions was assessed by 20 experts who evaluated the generated headlines. The findings indicate that the \textit{one-off intent expression} condition was, on average, capable of producing high-quality headlines without further iterations. This suggests that a simple textbox interface such as chatbot can adequately serve as an IUI for one-off intent expressions in fixed-scope content curation tasks similar to headline generation.

\subsection{Study 2 on Cross-Domain Analogy Generation: Intent Iteration for Atomic Creative Tasks}

\begin{figure}
\centering
\includegraphics[width=1\linewidth]{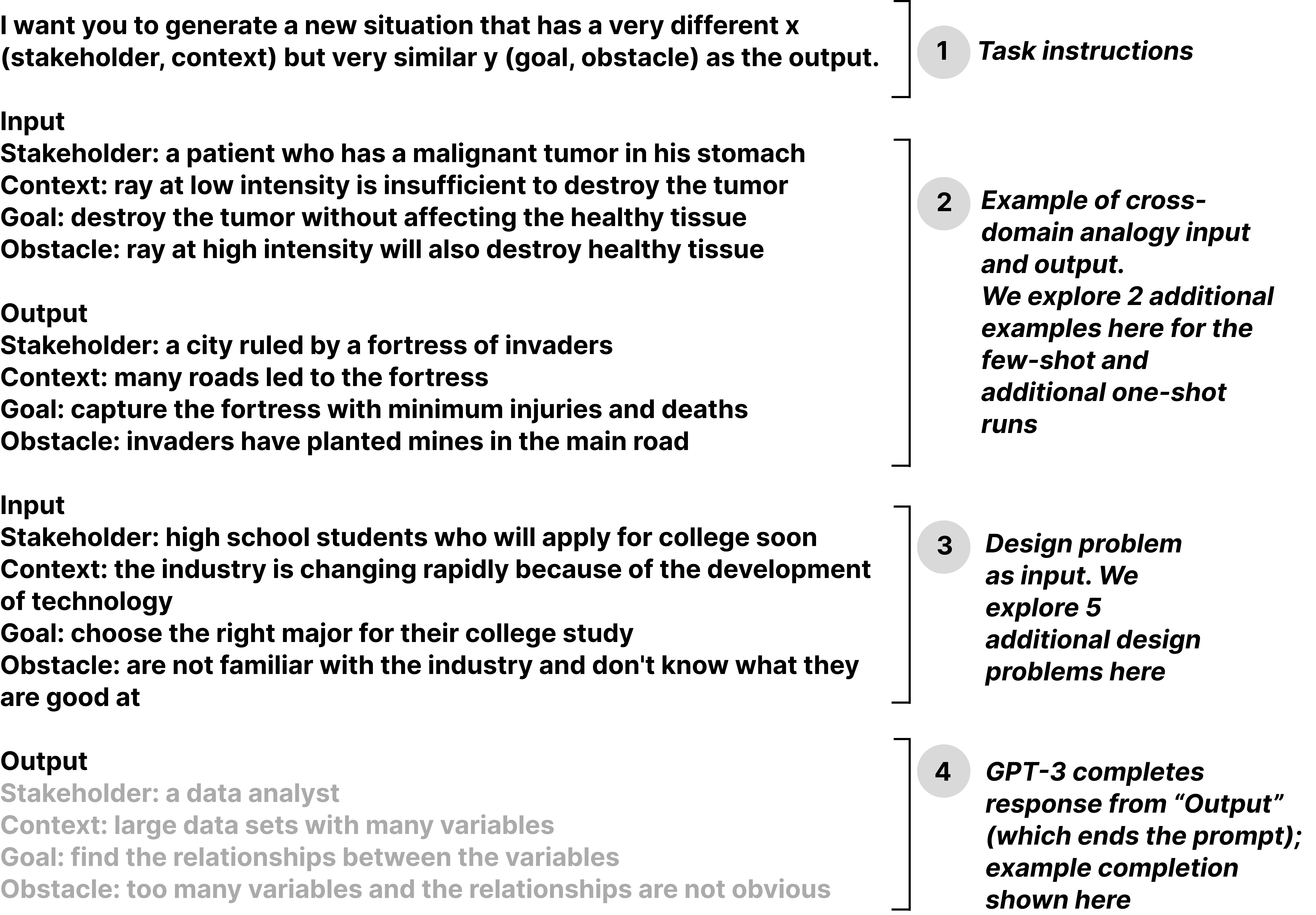}
\caption{Study 2 on atomic creative tasks - the prompt for the intent of generating cross-domain analogies based on Duncker and Lees' radiation problem \cite{dunckerProblemsolving1945}.}
\label{fig:prompt_structure}
\end{figure}

Transitioning from fixed-scope content curation tasks, our second study investigates atomic creative tasks, specifically focusing on cross-domain analogy generation. Unlike one-off intent expressions for fixed-scope content curation tasks, these tasks may require additional iterations to achieve high-quality outputs due to their open-ended nature. In our exploratory experiments, we evaluated the interaction between intent and AI in these tasks using prompts that guide Large Language Models (LLMs) with classical problems known for their analogical complexity, such as Duncker and Lees' radiation problem \cite{dunckerProblemsolving1945} (illustrated in Figure \ref{fig:prompt_structure}).

To evaluate the effectiveness of cross-domain analogies generated by LLM (GPT-3 \textit{text-davinci-002}), participants were tasked with using these analogies to reformulate original design problems \cite{ding2023fluid}. The findings were promising: the majority of AI-generated analogies were deemed useful, receiving a median helpfulness rating of 4 out of 5. Furthermore, these analogies facilitated significant changes in problem formulation in approximately 80\% of the cases. However, concerns were raised as up to 25\% of the outputs were considered potentially harmful due to the presence of unsettling content. These insights highlight the potential of intent-AI interaction in atomic creative tasks and simultaneously draw attention to the necessity for \textit{intent iteration} in the format of \textit{selection}, \textit{rating}, and \textit{post-editing} to address potential biases and ensure ethical and legal compliance.

\subsection{Study 3 (in progress) on Exploratory Visual Data Analysis: Intent Exploration, Iteration, and Sensemaking for Complex and Interdependent Tasks}

\begin{figure}
\centering
\includegraphics[width=0.78\linewidth]{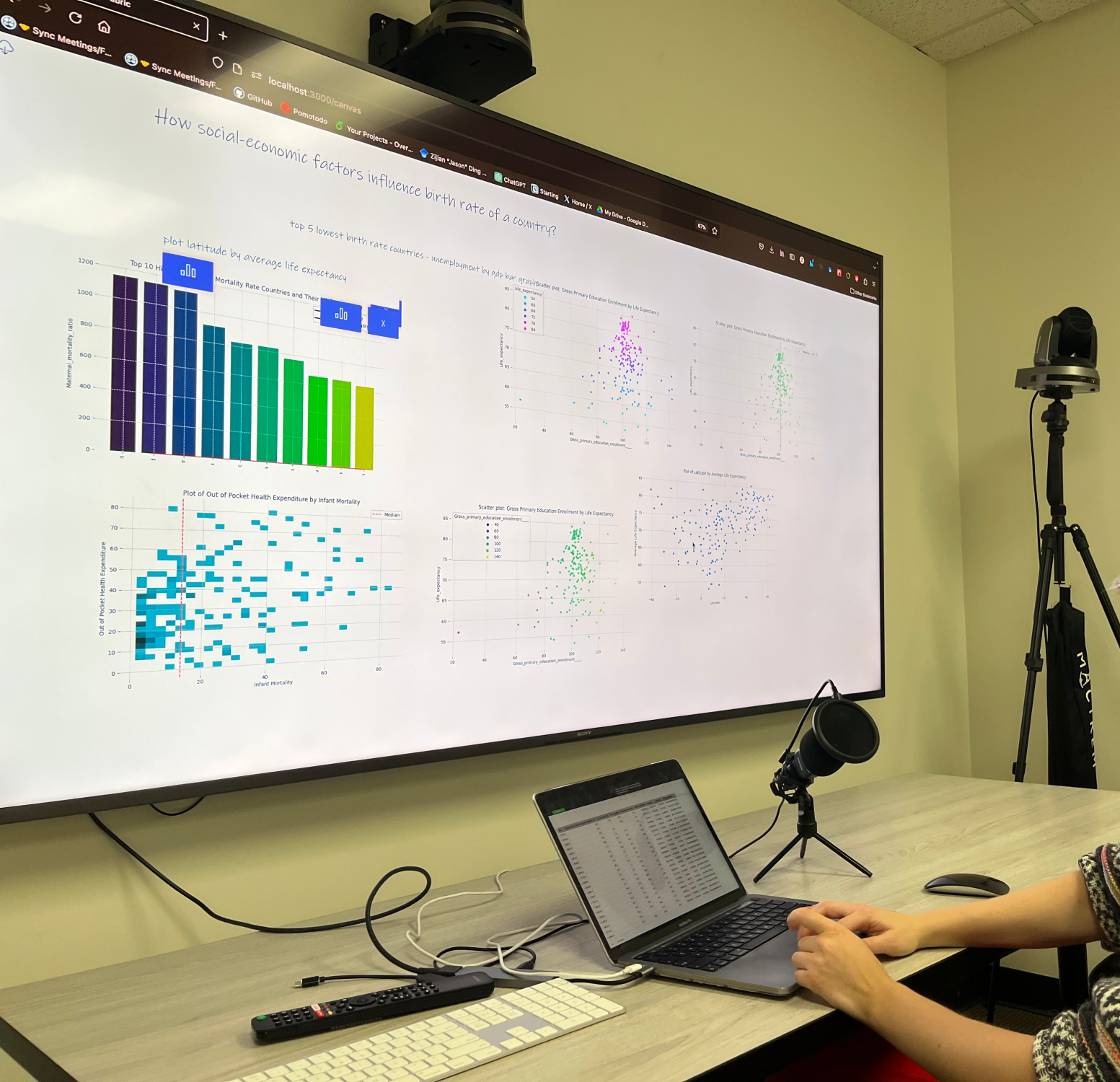}
\caption{Study 3 on exploratory visual data analysis - the participant explored and iterated on analysis intents in natural language, and organized generated visualization on a large display.}
\label{fig:setup}
\end{figure}

We extended our investigation into intent-AI interactions to complex and interdependent tasks, specifically focusing on exploratory visual data analysis. As a preliminary study, we developed a freeform canvas as Intent-based User Interface (IUI) where users can swiftly articulate their analytical intents using natural language prompts to generate and iteratively refine visualizations as "space to think" \cite{knudsenExploratoryStudyHow2012}. This process also includes the (re)organization of visualization results for sensemaking on a large display (illustrated in Figure \ref{fig:setup}).

To understand how users engage with this experimental interface, we conducted a user study involving 10 data analysts, ranging from novices to experienced professionals, complemented by semi-structured interviews. Initial findings revealed differences in interaction patterns between the two groups. Experienced data analysts predominantly engaged in iterative refinement of their visualizations, often using detailed, command-like expressions of intent, such as "Correlation plot of data - (+) correlation coefficient in red, (-) correlation coefficient in blue; include all variables that are continuous". This approach yielded high-quality outputs but limited the exploration of alternative analytical perspectives, potentially overlooking the exploratory nature of the task.

Conversely, novice analysts were more inclined to explore broadly, utilizing higher-level intent expressions like "What is the relationship between density and birth rate?" or "Can you tell me the trend?" However, the outputs generated from these prompts were of lower quality, indicating that novices might benefit from more structured guidance and scaffolding throughout their analytical endeavors.

These observations highlight the diverse needs of users at different levels of expertise when interacting with intent-driven AI tools in complex and interdependent tasks. They emphasize the necessity for IUI that can better support both broad exploratory intent expressions and detailed intent iterations. In the following studies, I will further explore the development of IUI for such tasks, focusing on providing experienced users with suggestions to engage more with high-level intents rather than solely focusing on iterations, and more structured guidance and scaffolds in IUI to aid novice users in achieving higher quality results.

\bibliographystyle{ACM-Reference-Format}
\bibliography{sample-base}

\end{document}